\documentclass{Interspeech}

\interspeechcameraready

\title{SEED: Speaker Embedding Enhancement Diffusion Model}

\author[affiliation={1}]{Kihyun}{Nam}
\author[affiliation={2}]{Jungwoo}{Heo}
\author[affiliation={3}]{Jee-weon}{Jung$^\dagger$}
\author[affiliation={4}]{Gangin}{Park}
\author[affiliation={1}]{Chaeyoung}{Jung}
\author[affiliation={2}]{Ha-Jin}{Yu} 
\author[affiliation={1}]{Joon Son}{Chung}

\affiliation{School of Electrical Engineering}{Korea Advanced Institute of Science and Technology}{South Korea}
\affiliation{School of Computer Science}{University of Seoul}{South Korea}
\affiliation{Language Technologies Institute}{Carnegie Mellon University}{United States}
\affiliation{Department of Electrical and Computer Engineering}{Seoul National University}{South Korea}

\email{nkh.mmai@kaist.ac.kr, jungwoo4021@gmail.com, jeeweonj@ieee.org, ssonpull519@snu.ac.kr, codud9914@kaist.ac.kr, hjyu@uos.ac.kr, joonson@kaist.ac.kr}


\usepackage{comment}
\usepackage{booktabs}
\usepackage{amsmath}
\usepackage{graphicx}
\usepackage{subcaption}
\usepackage{multirow}
\usepackage[subtle]{savetrees}
\usepackage{array}
\usepackage[table]{xcolor}
\usepackage{caption}

\newcommand{\newpara}[1]{\vspace{1mm}\noindent\textbf{#1}}

\makeatletter
\renewcommand{\section}{\@startsection{section}{1}{0pt}%
  {6pt} 
  {3pt} 
  {\normalfont\normalsize\bfseries}}  

\renewcommand{\subsection}{\@startsection{subsection}{2}{0pt}%
  {2pt} 
  {1pt} 
  {\normalfont\normalsize\bfseries}}  

\renewcommand{\subsubsection}{\@startsection{subsubsection}{3}{0pt}%
  {2pt} 
  {1pt} 
  {\normalfont\normalsize\itshape}}  
\makeatother

\begin{document}

\setlength{\abovedisplayskip}{2pt}  
\setlength{\belowdisplayskip}{2pt}  
\setlength{\abovedisplayshortskip}{1pt} 
\setlength{\belowdisplayshortskip}{1pt} 
\maketitle

\begin{abstract}
\vspace{-1mm}
A primary challenge when deploying speaker recognition systems in real-world applications is performance degradation caused by environmental mismatch. We propose a diffusion-based method that takes speaker embeddings extracted from a pre-trained speaker recognition model and generates refined embeddings. For training, our approach progressively adds Gaussian noise to both clean and noisy speaker embeddings extracted from clean and noisy speech, respectively, via forward process of a diffusion model, and then reconstructs them to clean embeddings in the reverse process. While inferencing, all embeddings are regenerated via diffusion process. Our method needs neither speaker label nor any modification to the existing speaker recognition pipeline. Experiments on evaluation sets simulating environment mismatch scenarios show that our method can improve recognition accuracy by up to 19.6\% over baseline models while retaining performance on conventional scenarios. We publish our code here\footnote{Official code: \href{https://github.com/kaistmm/seed-pytorch}{{https://github.com/kaistmm/seed-pytorch}}}.

\end{abstract}

\begingroup
  \renewcommand\thefootnote{}%
  \footnote{$^\dagger$Currently at Apple}
  \addtocounter{footnote}{-1}%
\endgroup

\vspace{-5mm}
\section{Introduction}
\vspace{-1mm}
Speaker recognition systems are widely used in various fields such as user authentication, security, and voice interfaces. However, differences in recording equipment, background noise, and other environmental factors in real-world conditions can introduce substantial acoustic discrepancies between utterances from the same speaker, degrading the recognition performance~\cite{campbell1997speaker, nam24b_interspeech}. Consequently, achieving robustness against such domain or environment mismatches has become a critical challenge for real-world applications. 

Such domain mismatch problems, including environment mismatch scenarios, ultimately widen the gap in the speaker embedding space and reduce the speaker identity similarity among different utterances from the same speaker. To mitigate this problem, a variety of methods have been proposed, including training with large-scale datasets~\cite{nagrani2020voxceleb, chung18b_interspeech} containing real-world noise, simulating diverse noise conditions through data augmentation~\cite{snyder2015musan, ko2017study}, and leveraging Disentangled Representation Learning (DRL)~\cite{nam24b_interspeech, wang2022disentangled, nam23_interspeech} techniques to separate and suppress task-irrelevant information. Although these approaches have shown some effectiveness, they often require complex network architectures, numerous hyperparameters, and extensive training resources.

\begin{figure*}
    \centering
    \includegraphics[width=0.9\linewidth]{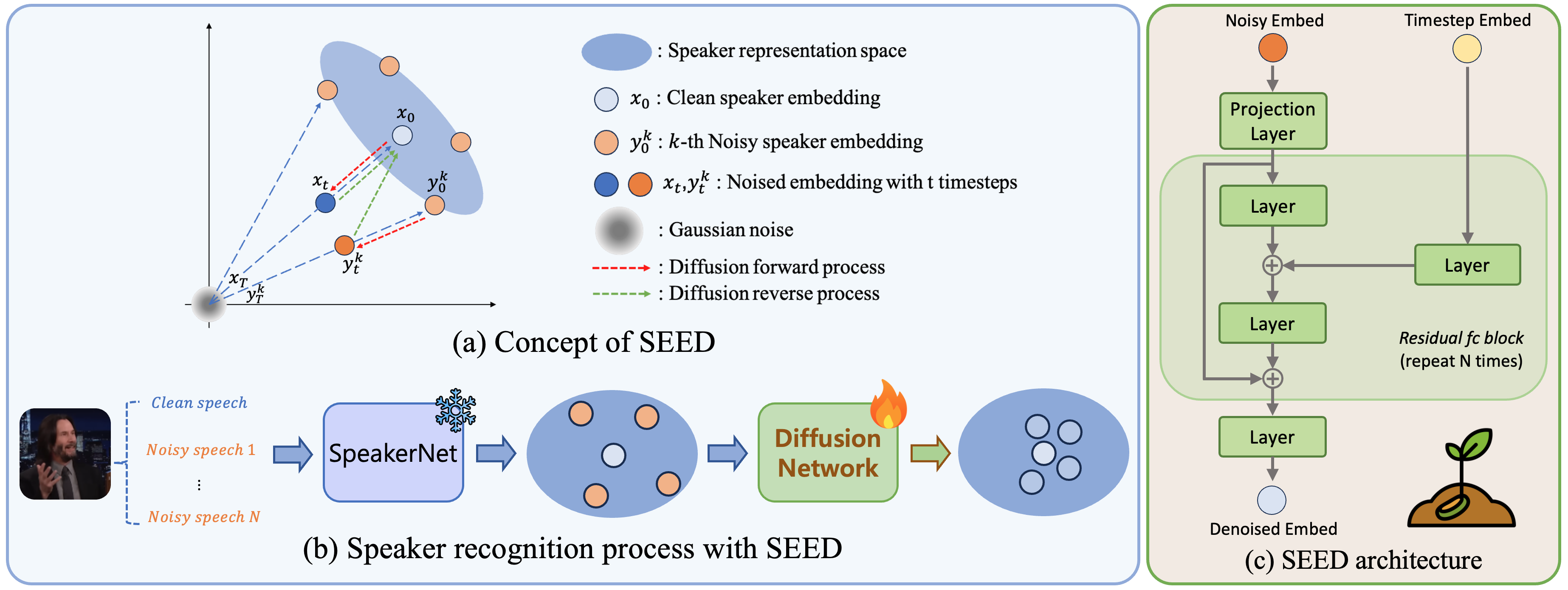}
    \vspace{-3mm}
    \caption{Illustration of Speaker Embedding Enhancement Diffusion (SEED) model. (a) explains the concept of our diffusion mechanism. (b) shows the whole training process of SEED. (c) illustrates the architecture of SEED.}
    \label{fig:main_figure}
    \vspace{-6mm}
\end{figure*}

Recently, Diffusion Probabilistic Model (DPM)~\cite{ho2020denoising, song2021denoising} has gained significant attention as a powerful generative model, capable of producing high-fidelity data in various domains, such as image generation and speech enhancement. The core principle of DPM is to incrementally inject noise (Forward process) into the data until it approximates a Gaussian distribution, and then gradually remove this noise (Reverse process) to recover the original data distribution. In the audio domain, DPM-based techniques have primarily been explored for front-end signal enhancement, where noisy waveforms or spectrograms are transformed into cleaner signals before being used by downstream tasks.  Although these studies effectively extend existing noise suppression and filtering methods, they generally operate on raw audio signals, offering no guarantee of directly resolving domain mismatch in the speaker embedding space. Thus, applying diffusion models at the embedding level, where speaker identity is explicitly encoded, remains largely unexplored, leaving open the question of whether front-end audio enhancement alone can ensure higher-quality speaker embeddings.

In this study, we propose a novel DPM-based approach that minimises the discrepancy between speaker embeddings extracted from clean and noisy speech samples. Our method is built on the assumption that, since both the clean embedding and the noisy embedding (obtained from noisy audio) inherently encapsulate the same speaker identity, their difference in the embedding space is not excessively large. By applying the forward DPM process, we progressively add Gaussian noise to both clean speaker embedding and noisy speaker embedding, causing their distributions to converge toward an almost identical Gaussian representation. Subsequently, the reverse process incrementally removes the noise, reconstructing the embeddings into a clean form.  In this process, even embeddings extracted under adverse conditions are refined into more clean representations, ensuring that utterances from the same speaker remain closely aligned in the embedding space, regardless of external recording factors. We refer to this DPM-based approach as the Speaker Embedding Enhancement Diffusion (SEED) model. To the best of our knowledge, SEED is the first attempt in speaker recognition to apply a diffusion model directly at the embedding level.

Our proposed SEED has several benefits. It requires no additional training or structural modifications to an already-trained speaker recognition model, making it versatile. It can be trained and applied on top of arbitrary speaker embeddings. Secondly, SEED does not rely on speaker labels; a corpus of clean speech where acoustic augmentation can be applied is enough --- there are millions of hours of such data, e.g., corpus designed for ASR or TTS. Lastly, SEED’s lightweight architecture, based on simple residual fully-connected (fc) blocks, demonstrates robust performance under environment mismatch condition, surpassing existing methods. Empirical evaluations across diverse speaker recognition systems and datasets show that SEED improves speaker identification accuracy by up to 19.6\% over baseline methods in mismatched conditions, underscoring the potential of embedding-level diffusion as a fundamental solution for real-world speaker recognition system deployments.

\vspace{-1mm}
\section{Related works}

\subsection{Diffusion Probabilistic Model}
\label{sec:dpm-related}

Diffusion probabilistic models (DPMs)~\cite{ho2020denoising, song2021denoising} have emerged as a powerful generative framework for high-fidelity data generation across various domains~\cite{ramesh2022hierarchical, rombach2022high, liu2023audioLDM}. These models operate through a forward process that gradually corrupts data with Gaussian noise, followed by a reverse process that reconstructs the original data distribution. However, diffusion models require numerous sampling steps, leading to slow inference. To address this, more efficient techniques, such as the Denoising Diffusion Implicit Model (DDIM)~\cite{song2021denoising}, have been introduced, utilising a non-Markovian process to accelerate sampling. In this work, we employ DDIM sampling to reduce the number of reverse steps while maintaining the quality of generated embeddings.

\subsection{Feature enhancement} 

\vspace{-1mm}
\newpara{Speech enhancement.} Speech enhancement primarily aims to reduce noise and reverberation. Traditional approaches rely on spectral subtraction~\cite{boll1979suppression}, Wiener filtering~\cite{lim1979enhancement}, or statistical methods~\cite{ephraim1984speech}. More recently, deep learning has enabled data-driven techniques, including DNN-based autoencoders~\cite{lu2013speech}, CNNs~\cite{pandey2019new}, GANs~\cite{pascual17_interspeech}, VAEs~\cite{leglaive2018variance}, and diffusion-based restoration~\cite{welker22_interspeech}. However, integrating these enhanced signals into downstream tasks often proves challenging: while intelligibility may improve, distortions can arise that compromise speaker-specific information~\cite{iwamoto22_interspeech, plchot2016audio}. Joint optimisation efforts~\cite{eskimez2018front, wu2021joint} highlight the difficulty of fully preserving speaker-relevant features in enhancement-based speaker recognition systems.

\vspace{-1mm}
\newpara{Speaker embedding enhancement.} Enhancing speaker embeddings to address domain mismatch has been extensively studied. Early methods based on i-vector and x-vector frameworks employed PLDA and length normalisation~\cite{garcia2011analysis}, but these often degrade under severe noise or unseen conditions. DRL aims to isolate speaker identity from nuisance factors~\cite{nam23_interspeech, nam24b_interspeech}, yet typically demands additional labeling and design of specialised training objectives. More recently, refining extracted speaker embeddings has shown promise for mitigating domain shifts; for instance, ~\cite{jung20_odyssey} proposed post-processing to reduce variability introduced by recording conditions. Our work proposes a diffusion-based approach that enhances speaker embeddings directly, requiring no speaker labels and allowing the use of clean speech datasets of various audio tasks. 

\vspace{-1mm}
\section{Proposed method}

This section describes the configuration of our proposed SEED, including the batch setup with audio augmentation and the speaker embedding extraction process. The concept, process, and model structure of our methodology are shown in Figure~\ref{fig:main_figure}.

\subsection{Batch configuration with multi-pair audio augmentation}
\label{sec:batch-config}
To simulate environment mismatch scenarios, we configure each mini-batch to contain pairs of \textit{clean} and \textit{noisy} audio data. Since the task involves transforming speaker embeddings extracted from noisy audio to match those from clean audio, it is essential to ensure that both embeddings share the same speaker identity and content. Noisy audio samples are generated through audio augmentation applied to one clean audio data \( a_{\text{clean}} \). Specifically, perturbations such as environmental noise and reverberation are introduced to simulate diverse real-world conditions. This process produces \( N \) noisy variants, $a_\text{noisy}^{\text{0}},  a_\text{noisy}^{\text{1}},  \dots, a_\text{noisy}^{\text{N-1}}$,
where each variant retains the same content as the clean audio but exhibits different acoustic characteristics. These variants allow the diffusion model to generalise by learning robust reconstruction paths in the reverse process. For the \( i \)-th mini-batch, the paired audio data can be represented as:
\[
(a_{\text{clean}}, \{ a_\text{noisy}^{\text{0}}, a_\text{noisy}^{\text{1}}, \dots, a_\text{noisy}^{\text{N-1}} \}).
\]

Details of audio augmentation techniques are further described in Section~4.

\subsection{Speaker embedding extractor}

Our method utilises a \textit{pre-trained speaker embedding network} that is not jointly optimised with SEED. This network extracts a \( D \)-dimensional speaker embedding vector from input audio \( a \), capturing essential speaker identity features. The speaker embedding extraction process can be expressed as:
\[
\mathbf{x}_0  = SpeakerNet(a_{clean}), \quad \mathbf{x}_0 \in \mathbb{R}^D,
\]
where \( \mathbf{x}_0 \) is the corresponding clean speaker embedding and $SpeakerNet$ denotes speaker embedding network. Similarly, embeddings extracted from input noisy audio \( a_{noisy}^k \)  are denoted by \( \mathbf{y}_0^{k} \), where \( k \) indicates the \( k \)-th noisy variant. The embedding network typically consists of three main components: a feature extraction module (e.g., STFT or convolutional layers), an encoder network, and a pooling mechanism that aggregates temporal features into a fixed-size vector. These embeddings are then passed to the diffusion model for further enhancement. More details of the speaker embedding network architecture will be presented in Section~\ref{sec:experiments}.

\subsection{Speaker Embedding Enhancement Diffusion}

Our SEED builds upon the Denoising Diffusion Probabilistic Model (DDPM) \cite{ho2020denoising}, 
extending it to handle both clean and noisy speaker embeddings in a unified framework. 
This section first revisits the basic DDPM formulation based on \emph{$\epsilon$-prediction} (estimates Gaussian noise), 
then illustrates how we adapt it to directly estimate a clean speaker embedding $\mathbf{x}_0$ from the noisy speaker embedding $\mathbf{y}_0$ via \emph{sample-prediction} (estimates target sample directly) formulation.

\subsubsection{Basic DDPM Formulation}
\label{sec:basicddpm}
Let $\mathbf{x}_0 \sim q(\mathbf{x}_0)$ be the original data (in our case, a speaker embedding). 
A forward diffusion process $q(\mathbf{x}_t \mid \mathbf{x}_0)$ progressively corrupts $\mathbf{x}_0$ 
with Gaussian noise over $t \in \{1,\dots,T\}$ steps, defined as:
\begin{equation}
    q(\mathbf{x}_t | \mathbf{x}_0) = \mathcal{N}(\mathbf{x}_t; \sqrt{\bar{\alpha}_t} \mathbf{x}_0, (1 - \bar{\alpha}_t) \mathbf{I}),
\label{eq:forward-dist}
\end{equation}
where $\alpha_t \in (0,1)$ is a noise scheduling parameter, $\overline{\alpha}_t = \prod_{\tau=1}^t \alpha_\tau$, 
and $\mathbf{I}$ is the identity matrix. 
To estimate $q(\mathbf{x}_{t-1} | \mathbf{x}_t)$, the parameterised reverse process $p_\theta(\mathbf{x}_{t-1}\mid \mathbf{x}_t)$ aims to \emph{denoise} $\mathbf{x}_t$ 
back to $\mathbf{x}_{t-1}$:
\begin{equation}
\label{eq:reverse-dist}
    p_\theta(\mathbf{x}_{t-1} \mid \mathbf{x}_t) 
    = \mathcal{N}\Bigl(\mathbf{x}_{t-1};\,\mu_\theta(\mathbf{x}_t,t),\,\sigma_t^2\,\mathbf{I}\Bigr),
\end{equation}
where a neural network parameterised by $\theta$ predicts the mean $\mu_\theta(\mathbf{x}_t,t)$,
and $\sigma_t^2 = 1-\alpha_t$ follows the same noise schedule. 
With standard DDPM, it is typically trained $\theta$ by minimising a variational lower bound on the data likelihood. As mentioned Section~\ref{sec:dpm-related}, we adopt an implicit sampler \cite{song2021denoising} (DDIM).

\subsubsection{Diffusion Process of SEED}
\label{sec:reverse-eps}
In forward process (the red arrow dash lines in Figure~\ref{fig:main_figure}-a), we interpret $\mathbf{x}_0$ as a \emph{clean} speaker embedding 
and $\mathbf{y}_0$ as its \emph{noisy} counterpart. 
Applying Eq.~\eqref{eq:forward-dist}, 
we can sample a corrupted state $\mathbf{x}_t$ via
\begin{equation}
\label{eq:forward-x}
    \mathbf{x}_t 
    = \sqrt{\overline{\alpha}_t}\,\mathbf{x}_0 
    + \sqrt{1 - \overline{\alpha}_t}\,\boldsymbol{\epsilon}, 
    \quad 
    \boldsymbol{\epsilon}\sim\mathcal{N}(0,\mathbf{I}),
\end{equation}
and similarly, a corresponding corrupted \emph{noisy} embedding $\mathbf{y}_t$ can be written as 
\begin{equation}
\label{eq:forward-y}
    \mathbf{y}_t 
    = \sqrt{\overline{\alpha}_t}\,\mathbf{y}_0 
    + \sqrt{1 - \overline{\alpha}_t}\,\boldsymbol{\epsilon}.
\end{equation}
Thus, $\mathbf{x}_t$ and $\mathbf{y}_t$ are Gaussian corrupted embeddings, each from $\mathbf{x}_0$ and $\mathbf{y}_0$, reparameterised with $\boldsymbol{\epsilon}$.

In reverse process (the green arrow dash lines in Figure~\ref{fig:main_figure}-a), based on Eq.~\eqref{eq:reverse-dist}, 
the reverse step $p_\theta(\mathbf{x}_{t-1}\mid \mathbf{x}_t)$ 
is instantiated by a network $\epsilon_\theta$ that does \emph{$\epsilon$-prediction} which estimates $\boldsymbol{\epsilon}$, yielding the reconstruction $(\mathbf{x}_t \to \mathbf{x}_0)$ as:
\begin{equation}
\label{eq:ddpm-x0}
    \hat{\mathbf{x}}_0(\mathbf{x}_t, t)
    = \frac{1}{\sqrt{\overline{\alpha}_t}}
      \Bigl(\mathbf{x}_t - \sqrt{1-\overline{\alpha}_t}\,\epsilon_\theta(\mathbf{x}_t,t)\Bigr).
\end{equation}
This reconstructs the clean embedding $\mathbf{x}_0$ from $\mathbf{x}_t$.  
However, standard DDPM addresses only a single distribution (either $q(\mathbf{x}_0)$ or $q(\mathbf{y}_0)$), 
whereas SEED also seeks reconstruction of $\mathbf{y}_0$ from $\mathbf{y}_t$ to be closer to $\mathbf{x}_0$. 
To achieve this, we consider Eq.~\eqref{eq:ddpm-x0} for $\mathbf{y}_0$ as:
\begin{equation}
\label{eq:ddpm-y0}
    \hat{\mathbf{y}}_0(\mathbf{y}_t,t)
    = \frac{1}{\sqrt{\overline{\alpha}_t}}
      \Bigl(\mathbf{y}_t - \sqrt{\,1-\overline{\alpha}_t}\,\epsilon_\theta(\mathbf{y}_t,t)\Bigr)
    \approx \mathbf{x}_0.
\end{equation}
Unlike the standard reverse process $(\mathbf{x}_t \to \mathbf{x}_0)$ or $(\mathbf{y}_t \to \mathbf{y}_0)$, this cross-reconstruction $(\mathbf{y}_t \to \mathbf{x}_0)$ acts as \emph{noise removal} in the speaker embedding space.

\subsubsection{Training Objective}
\label{sec:seed_obj}

SEED jointly considers two objectives: (i) learning distribution of clean speaker embedding $\mathbf{x}_0$ for \emph{high-quality} sampling, and (ii) learning the reverse process of $(\mathbf{y}_t \to \mathbf{x}_0)$ for embedding enhancement of Eq.~\eqref{eq:ddpm-y0} via direct optimisation of 
\begin{equation}
    ||\mathbf{x}_0 - \hat{\mathbf{y}}_0(\mathbf{y}_t,t)||,
\end{equation}
which introduces an additional scaled noise term as

\vspace{-3mm}
\begin{equation}
\label{eq:scaled_noise}
\begin{aligned}
||\mathbf{x}_0 - \hat{\mathbf{y}}_0(\mathbf{y}_t,t)||
&=
\Bigl|\Bigl| \frac{1}{\sqrt{\overline{\alpha}_t}}
\Bigl(\mathbf{x}_t - \sqrt{\,1-\overline{\alpha}_t}\,\boldsymbol{\epsilon}\Bigr)
\\
&\quad -
\frac{1}{\sqrt{\overline{\alpha}_t}}
\Bigl(\mathbf{y}_t - \sqrt{\,1-\overline{\alpha}_t}\,\epsilon_\theta(\mathbf{y}_t,t)\Bigr)
\Bigr|\Bigr|
\end{aligned}
\end{equation} in terms of \emph{$\epsilon$-prediction}. Rearranging the terms within the norm reveals
\[ \mathbf{x}_t - \mathbf{y}_t  = \sqrt{\overline{\alpha}_t}\,\bigl(\mathbf{x}_0 - \mathbf{y}_0\bigr), \]
which acts as an additional \emph{scaled noise} factor. If $||\mathbf{x}_0 - \mathbf{y}_0||$ is large, this scaled noise term may destabilise the training of the reverse process. However, under our assumption that $\mathbf{x}_0$ and $\mathbf{y}_0$ come from the same utterance of the same speaker, we expect the scaled noise to be moderate and the model learns to regress the scaled noise for each timestep accordingly.  Furthermore, the \textit{multi-pair audio augmentation} in Section~\ref{sec:batch-config} ensures the model encounters diverse noisy variants $\{\mathbf{y}_0^{k}\}$ around each $\mathbf{x}_0$, improving generalisation to various scaled noise patterns.

Finally, we propose a loss function that can implicitly optimise the scaled noise and epsilon terms via \emph{sample-prediction}:
\begin{equation}
L_{\mathrm{SEED}}
= 
\mathbb{E}\Bigl[
||\mathbf{x}_0 - f_{\theta}(\mathbf{x}_t,t)|| 
+
\sum_{k=0}^{N-1}
|| \mathbf{x}_0 - f_{\theta}(\mathbf{y}_t^{k},t)||
\Bigr],
\label{eq:loss-seed}
\end{equation}
Here, $f_{\theta}$ denotes the trainable SEED network. Specifically, $f_{\theta}(\mathbf{x}_t,t)$ and $f_{\theta}(\mathbf{y}_t^{k},t)$ directly predict $\hat{\mathbf{x}}_0$ and $\hat{\mathbf{y}}_0^{k}$, respectively. As a result, SEED can (i) ensure high-quality speaker embedding generation for clean samples and (ii) remove noise from $\mathbf{y}_t^{k}$ effectively, all within a unified diffusion framework.

\begin{table*}[t]
\centering
\caption{
EER and minDCF Results on five evaluation sets. \textbf{boldface score} indicates that the model achieved the highest performance among the comparison groups on the same baseline and evaluation set. $\rightarrow$ means passing the output from the left model to the right model. $+$ means scratch training the baseline network with additional strategies or networks on the right. $\uparrow$ indicates performance improvement over baseline.}
\vspace{-2mm}

\resizebox{0.7\linewidth}{!}{
\begin{tabular}{lcccccccccc}
\toprule
& \multicolumn{4}{c}{\textbf{Environmental mismatch set}}
& \multicolumn{6}{c}{\textbf{Generalization set}} \\
\cmidrule(lr){2-5}\cmidrule(lr){6-11}
& \multicolumn{2}{c}{\bf VoxSRC23\phantom{$\uparrow$}} 
& \multicolumn{2}{c}{\bf VC-Mix\phantom{$\uparrow$}} 
& \multicolumn{2}{c}{\bf Vox1-O\phantom{$\uparrow$}} 
& \multicolumn{2}{c}{\bf Vox1-E\phantom{$\uparrow$}} 
& \multicolumn{2}{c}{\bf Vox1-H\phantom{$\uparrow$}} \\
\cmidrule(lr){2-3}\cmidrule(lr){4-5}
\cmidrule(lr){6-7}\cmidrule(lr){8-9}\cmidrule(lr){10-11}
\bf Model
& \bf EER\phantom{$\uparrow$} & \bf minDCF\phantom{$\uparrow$}
& \bf EER\phantom{$\uparrow$} & \bf minDCF\phantom{$\uparrow$}
& \bf EER\phantom{$\uparrow$} & \bf minDCF\phantom{$\uparrow$}
& \bf EER\phantom{$\uparrow$} & \bf minDCF\phantom{$\uparrow$}
& \bf EER\phantom{$\uparrow$} & \bf minDCF\phantom{$\uparrow$} \\
\midrule

\multicolumn{11}{c}{\textbf{Baseline}} \\
\midrule
\text{ResNet34~\cite{kwon2021ins}}
& 5.50\phantom{$\uparrow$} & 0.308\phantom{$\uparrow$}
& 3.07\phantom{$\uparrow$} & 0.245\phantom{$\uparrow$}
& 0.88\phantom{$\uparrow$} & 0.079\phantom{$\uparrow$}
& 1.07\phantom{$\uparrow$} & 0.076\phantom{$\uparrow$}
& 2.21\phantom{$\uparrow$} & 0.147\phantom{$\uparrow$}
\\

\text{ECAPA-TDNN~\cite{desplanques2020ecapa}}
& 5.93\phantom{$\uparrow$} & 0.335\phantom{$\uparrow$}
& 2.96\phantom{$\uparrow$} & 0.261\phantom{$\uparrow$}
& 0.90\phantom{$\uparrow$} & 0.064\phantom{$\uparrow$}
& 1.16\phantom{$\uparrow$} & 0.081\phantom{$\uparrow$}
& 2.38\phantom{$\uparrow$} & 0.152\phantom{$\uparrow$}
\\

\text{WavLM-ECAPA~\cite{chen2022wavlm}}
& 5.07\phantom{$\uparrow$}  & 0.283\phantom{$\uparrow$}  

& 2.32\phantom{$\uparrow$}  & 0.195\phantom{$\uparrow$} 
& 0.83\phantom{$\uparrow$}  & 0.063\phantom{$\uparrow$} 
& 0.98\phantom{$\uparrow$}  & 0.058\phantom{$\uparrow$} 
& 1.92\phantom{$\uparrow$}  & 0.115\phantom{$\uparrow$} 
\\
\midrule

\multicolumn{11}{c}{\textbf{Disentangled Representation Learning}} \\
\midrule
\text{ResNet34 + DRL~\cite{nam24b_interspeech}}
& 5.35$\uparrow$ & 0.306$\uparrow$
& 2.58$\uparrow$ & 0.223$\uparrow$
& \textbf{0.86$\uparrow$} & \textbf{0.068$\uparrow$}
& 1.10\phantom{$\uparrow$} & 0.078\phantom{$\uparrow$}
& \textbf{2.20$\uparrow$} & \textbf{0.142$\uparrow$}
\\

\text{ECAPA-TDNN + DRL~\cite{nam24b_interspeech}}   
& 5.81$\uparrow$ & 0.325$\uparrow$
& 2.43$\uparrow$ & \textbf{0.212$\uparrow$}
& \textbf{0.82$\uparrow$} & 0.067\phantom{$\uparrow$}
& 1.16\phantom{$\uparrow$} & \textbf{0.080$\uparrow$}
& 2.38\phantom{$\uparrow$} & 0.156\phantom{$\uparrow$}
\\
\midrule

\multicolumn{11}{c}{\textbf{Embedding Enhancement Diffusion}} \\
\midrule
\text{ResNet34~\cite{kwon2021ins} $\rightarrow$ SEED}
& \textbf{5.31$\uparrow$} & \textbf{0.303$\uparrow$}
& \textbf{2.56$\uparrow$} & \textbf{0.219$\uparrow$}
& 0.87$\uparrow$  & 0.079\phantom{$\uparrow$}
& 1.09\phantom{$\uparrow$} & 0.076\phantom{$\uparrow$}
& 2.23\phantom{$\uparrow$} & 0.149\phantom{$\uparrow$}
\\

\text{ECAPA-TDNN~\cite{desplanques2020ecapa} $\rightarrow$ SEED}
& \textbf{5.53$\uparrow$} & \textbf{0.309$\uparrow$}
& \textbf{2.38$\uparrow$} & 0.227$\uparrow$
& 0.86$\uparrow$ & 0.069\phantom{$\uparrow$}
& 1.20\phantom{$\uparrow$} & 0.083\phantom{$\uparrow$}
& \textbf{2.30$\uparrow$} & \textbf{0.148$\uparrow$}
\\

\text{WavLM-ECAPA~\cite{chen2022wavlm}$\rightarrow$ SEED}
& \textbf{4.94}$\uparrow$ & \textbf{0.280}$\uparrow$ 
& \textbf{2.29}$\uparrow$ & 0.200\phantom{$\uparrow$} 
& \textbf{0.81}$\uparrow$ & 0.065\phantom{$\uparrow$} 
& \textbf{0.97}$\uparrow$  & 0.058\phantom{$\uparrow$} 
& 1.92\phantom{$\uparrow$} & 0.115\phantom{$\uparrow$}
\\

\bottomrule
\end{tabular}
}
\vspace{-3mm}
\label{tab:merged_five_sets}
\end{table*}
\begin{table*}[ht!]
\centering

\caption{Performance comparison between conventional audio enhancement-based approach and our embedding enhancement diffusion-based approach. $\rightarrow$ means passing the output from the left model to the right model. 
\textbf{Params (M)} denotes the number of model parameters in millions, \textbf{RTF} (Real-Time Factor) indicates the processing speed relative to real-time for end-to-end process, and \textbf{Memory (GiB)} represents the memory consumption during inference. \textbf{RTF} and \textbf{Memory} measurements are based on 4-second audio size. We report  the Equal Error Rate (EER) for \textbf{VC-Mix} and \textbf{Vox1-O} evaluation set.}

\vspace{-2mm}

\label{tab:enhancement_results}
\resizebox{0.6\linewidth}{!}{
\begin{tabular}{lccccc}
\toprule
\textbf{Model} & \textbf{Params (M)} & \textbf{RTF} & \textbf{Memory (GiB)} & \textbf{VC-Mix} & \textbf{Vox1-O} \\
\midrule
SE~\cite{defossez20_interspeech} $\rightarrow$ ResNet34~\cite{kwon2021ins} & 44.6 & 0.0041 & 0.88 & 5.14 & 1.21 \\
SE~\cite{defossez20_interspeech}  $\rightarrow$ ECAPA-TDNN~\cite{desplanques2020ecapa}  & 46.6 & 0.0038 & 0.88 & 4.63 & 1.17 \\
SE~\cite{defossez20_interspeech}  $\rightarrow$ WavLM-ECAPA~\cite{chen2022wavlm}  & 137.7 & 0.0069 & 2.89 & 31.29 & 30.67 \\
 SE~\cite{defossez20_interspeech}  $\rightarrow$ WavLM-ECAPA~\cite{chen2022wavlm} (fine-tuned) & 137.7 & 0.0069 & 2.89 & 2.24 & 0.81 \\
\midrule
\text{ResNet34~\cite{kwon2021ins} $\rightarrow$ SEED}  & 20.8 & 0.0022 & 0.59 & 2.56 & 0.87 \\
\text{ECAPA-TDNN~\cite{desplanques2020ecapa} $\rightarrow$ SEED}  & 18.8 & 0.0020 & 0.59 &  2.38 & 0.86 \\
\text{WavLM-ECAPA~\cite{chen2022wavlm} $\rightarrow$ SEED}  & 105.6 & 0.0030 & 2.66 & 2.29 & 0.81 \\

\bottomrule
\end{tabular}}
\vspace{-6mm}
\end{table*}

\vspace{-1mm}
\section{Experiments}
\label{sec:experiments}
\vspace{-1mm}

\subsection{Model Configuration}

\subsubsection{Speaker embedding network}

We employ three pre-trained speaker embedding networks. For the spectrogram-based speaker network, we selected `\textbf{H / ASP}' version of \textbf{ResNet34}~\cite{kwon2021ins} and \textbf{ECAPA-TDNN}~\cite{desplanques2020ecapa}. For the raw-waveform-based speaker network, we chose ECAPA-TDNN network combined with `\textbf{WavLM Base+}' version of WavLM~\cite{chen2022wavlm}, named \textbf{WavLM-ECAPA}. The output speaker embedding dimensions of the ResNet34, ECAPA-TDNN, and WavLM-ECAPA models are 512, 256, and 192, respectively.

\subsubsection{Speaker Embedding Enhancement Diffusion (SEED)} The SEED model adopts a representation diffusion model (RDM)~\cite{li2024return} network composed of residual fc blocks. Each residual fc block consists of an input layer, a timestep embedding projection layer, and an output layer, where each layer includes a LayerNorm~\cite{lei2016layer}, a SiLU~\cite{elfwing2018sigmoid}, and a linear layer. RDM~\cite{li2024return} network is illustrated in Figure \ref{fig:main_figure}-c. For all baseline speaker recognition models, we use three residual fc blocks without the optional context layer, following the same configuration as~\cite{li2024return}. Additionally, the hidden unit size of each layer within the fc blocks is set to twice the dimension of the speaker embedding used in each baseline model.

\subsection{Implementation details}
\vspace{-1mm}

\newpara{Datasets.} For training, we use approximately 1,000 hours of clean speech data, comprising 460 hours from LibriTTS-R~\cite{koizumi23_interspeech} and 577 hours from Libri-Light~\cite{kahn2020libri}. Notably, \textbf{no speaker labels are required}, as our training only depends on clean speech data. For evaluation, we employ five evaluation protocols: the three VoxCeleb1 variants (Vox1-O, Vox1-E, and Vox1-H) to evaluate generalisation performance, and VoxSRC23~\cite{huh2024vox} along with VC-Mix~\cite{heo2023rethinking} to evaluate environment robustness, as these latter datasets reflect environment mismatch scenarios.

\vspace{-1mm}
\newpara{Multi-pair audio augmentation.} For multi-pair audio augmentation, we generate three types of noisy speech from each clean utterance by applying reverberations from a simulated RIR dataset \cite{ko2017study} and music and background noises from MUSAN \cite{snyder2015musan}. The SNR is randomly selected within 0--15\,dB for background noise and 5--15\,dB for music noise. Consequently, each utterance forms a set of four paired samples (one clean and three noisy variants).

\vspace{-1mm}
\newpara{Training \& Inference.} All of our experiments utilise the PyTorch framework~\cite{paszke2019pytorch} together with the open-source \texttt{voxceleb\_trainer}\footnote{\url{https://github.com/clovaai/voxceleb_trainer}}. We select AdamW Optimizer~\cite{loshchilov2018decoupled} with an initial learning rate of 0.0005. Our implementation is conducted on a single NVIDIA RTX 4090 GPU with 24 GB memory. Training takes around 60 epochs. For the diffusion process, we use a scaled linear noise schedule with 1,000 training timesteps $T$ in the DDIM sampler~\cite{song2021denoising} from \texttt{diffusers}\footnote{\url{https://github.com/huggingface/diffusers}}. 

In the inference step, we assume that all input speaker embeddings will be $\mathbf{y}_{t}$ and regenerate them via the diffusion process of SEED. Unlike typical DDIM sampling, our \emph{sample-prediction} approach directly generates the cleansed output embedding from $\mathbf{y}_{t}$ in a single-step. We set the timestep \(t\) to 50 for single-step sampling. Empirically, we observed negligible performance differences compared to multi-step sampling. For WavLM-ECAPA, we apply a feature ensemble technique where the original speaker embedding and SEED's output embedding are summed to obtain the final speaker embedding.

\vspace{-1mm}
\newpara{Evaluation.}
We report two metrics: (1) Equal Error Rate (EER), where false accept and false reject rates converge, and (2) minimum Detection Cost Function (minDCF) from NIST SRE \cite{omid19nsre}, using \(C_{miss} = 1, C_{fa} = 1\), and \(P_{target} = 0.05\). For ResNet34 and ECAPA-TDNN, each utterance is split into ten 4-second segments, and all segment pairs are evaluated to yield an averaged similarity score. By contrast, WavLM-ECAPA uses the entire utterance. We follow the scoring protocols described in \cite{kwon2021ins, chen2022wavlm}.

\vspace{-2mm}
\section{Results}
\vspace{-2mm}
In this section, we summarise and analyse our experimental results. To assess both environment-robustness and generalisation performance across diverse speaker recognition systems, we compare three baselines, an adversarial DRL-based method~\cite{nam24b_interspeech} designed to remove environmental factors from the speaker representation, and our proposed method (SEED). These results are presented in Table~\ref{tab:merged_five_sets}.  Moreover, we evaluate the practical benefits of our approach compared to an existing audio enhancement-based approach~\cite{defossez20_interspeech}, demonstrating the efficacy of SEED under real-world constraints in Table~\ref{tab:enhancement_results}.

\vspace{-1mm}
\newpara{Environment-robustness performance.} As shown in Table~\ref{tab:merged_five_sets}, integrating our SEED framework with each baseline model yields up to a 19.6\% improvement on the environmental mismatch evaluation sets. Although a DRL approach~\cite{nam24b_interspeech} similarly achieves notable gains, it requires additional speaker and environment labels and more complex training procedures. In contrast, SEED requires no speaker labels and can be trained on a smaller training dataset, yet delivers competitive or superior outcomes. For example, on the VoxSRC23 set, applying SEED to ECAPA-TDNN outperforms the method in~\cite{nam24b_interspeech} by nearly 4.8\%. Furthermore, in the VC-Mix set, ECAPA-TDNN with SEED almost matches the performance of the WavLM baseline, which is pre-trained on a large datasets and diverse learning techniques. These results demonstrate that SEED effectively eliminates residual environmental information hidden in the speaker representation space, thereby improving the latent capacity of the speaker recognition system. 

\vspace{-1mm}
\newpara{Practical and high-fidelity speaker representation generation.} As shown in Table~\ref{tab:merged_five_sets}, our proposed method not only maintains remarkable generalisation performances on generalisation sets---achieving performance that is slightly higher or nearly indistinguishable from the baselines---but also delivers superior improvements on environment mismatch evaluation sets. This indicates that our approach consistently produces high-quality speaker embeddings. In other words, the diffusion-based paradigm demonstrates substantial potential for high-fidelity speaker representation generation at the embedding level.

Table~\ref{tab:enhancement_results} compares a conventional waveform-domain audio enhancement method with SEED. Despite requiring significantly fewer parameters, lower memory usage, and a reduced real-time factor, SEED outperforms most baselines on both VC-Mix and Vox1-O benchmarks. Notably, WavLM-ECAPA, which directly processes raw waveforms, suffers significant performance degradation unless it undergoes additional fine-tuning, whereas SEED provides both computational efficiency and robustness without any extra training. Consequently, it offers a compelling alternative to resource-intensive enhancement-based speaker recognition systems.

\vspace{-2mm}
\section{Conclusion}
\vspace{-2mm}

SEED leverages diffusion models for robust and generalised speaker embedding enhancement without requiring speaker labels or complex training. It seamlessly integrates with pre-trained speaker recognition models, enabling immediate deployment in speaker recognition systems for real-world applications. However, since SEED learns the gap between clean and noisy embeddings via the DDPM mechanism, some training instability may occur. To address this, future work will explicitly model the domain mismatch gap, ensuring stable generation even in extreme conditions and broadening SEED’s applicability to diverse speech tasks.

\clearpage

\section{Acknowledgements}
This work was supported by IITP grant funded by the Korea government (MSIT) (RS-2024-
00457882, National AI Research Lab Project).

\bibliographystyle{IEEEtran}
\bibliography{shortstrings,mybib}

\begin{thebibliography}{10}
\providecommand{\url}[1]{#1}
\csname url@samestyle\endcsname
\providecommand{\newblock}{\relax}
\providecommand{\bibinfo}[2]{#2}
\providecommand{\BIBentrySTDinterwordspacing}{\spaceskip=0pt\relax}
\providecommand{\BIBentryALTinterwordstretchfactor}{4}
\providecommand{\BIBentryALTinterwordspacing}{\spaceskip=\fontdimen2\font plus
\BIBentryALTinterwordstretchfactor\fontdimen3\font minus \fontdimen4\font\relax}
\providecommand{\BIBforeignlanguage}[2]{{%
\expandafter\ifx\csname l@#1\endcsname\relax
\typeout{** WARNING: IEEEtran.bst: No hyphenation pattern has been}%
\typeout{** loaded for the language `#1'. Using the pattern for}%
\typeout{** the default language instead.}%
\else
\language=\csname l@#1\endcsname
\fi
#2}}
\providecommand{\BIBdecl}{\relax}
\BIBdecl

\bibitem{campbell1997speaker}
J.~P. Campbell, ``Speaker recognition: A tutorial,'' \emph{Proceedings of the IEEE}, vol.~85, no.~9, pp. 1437--1462, 1997.

\bibitem{nam24b_interspeech}
K.~Nam, H.-S. Heo, J.~weon Jung, and J.~Chung, ``Disentangled representation learning for environment-agnostic speaker recognition,'' in \emph{Proc. Interspeech}, 2024, pp. 2130--2134.

\bibitem{nagrani2020voxceleb}
A.~Nagrani, J.~S. Chung, W.~Xie, and A.~Zisserman, ``Voxceleb: Large-scale speaker verification in the wild,'' \emph{Computer Speech \& Language}, vol.~60, p. 101027, 2020.

\bibitem{chung18b_interspeech}
J.~S. Chung, A.~Nagrani, and A.~Zisserman, ``{VoxCeleb2: Deep Speaker Recognition},'' in \emph{Proc. Interspeech}, 2018, pp. 1086--1090.

\bibitem{snyder2015musan}
D.~Snyder, G.~Chen, and D.~Povey, ``Musan: A music, speech, and noise corpus,'' \emph{arXiv preprint arXiv:1510.08484}, 2015.

\bibitem{ko2017study}
T.~Ko, V.~Peddinti, D.~Povey, M.~L. Seltzer, and S.~Khudanpur, ``A study on data augmentation of reverberant speech for robust speech recognition,'' in \emph{Proc. ICASSP}.\hskip 1em plus 0.5em minus 0.4em\relax IEEE, 2017, pp. 5220--5224.

\bibitem{wang2022disentangled}
X.~Wang, H.~Chen, S.~Tang, Z.~Wu, and W.~Zhu, ``Disentangled representation learning,'' \emph{arXiv preprint arXiv:2211.11695}, 2022.

\bibitem{nam23_interspeech}
K.~Nam, Y.~Kim, J.~Huh, H.-S. Heo, J.~weon Jung, and J.~S. Chung, ``Disentangled representation learning for multilingual speaker recognition,'' in \emph{Proc. Interspeech}, 2023, pp. 5316--5320.

\bibitem{ho2020denoising}
J.~Ho, A.~Jain, and P.~Abbeel, ``Denoising diffusion probabilistic models,'' \emph{Advances in neural information processing systems}, vol.~33, pp. 6840--6851, 2020.

\bibitem{song2021denoising}
\BIBentryALTinterwordspacing
J.~Song, C.~Meng, and S.~Ermon, ``Denoising diffusion implicit models,'' in \emph{International Conference on Learning Representations}, 2021. [Online]. Available: \url{https://openreview.net/forum?id=St1giarCHLP}
\BIBentrySTDinterwordspacing

\bibitem{ramesh2022hierarchical}
A.~Ramesh, P.~Dhariwal, A.~Nichol, C.~Chu, and M.~Chen, ``Hierarchical text-conditional image generation with clip latents,'' \emph{arXiv preprint arXiv:2204.06125}, vol.~1, no.~2, p.~3, 2022.

\bibitem{rombach2022high}
R.~Rombach, A.~Blattmann, D.~Lorenz, P.~Esser, and B.~Ommer, ``High-resolution image synthesis with latent diffusion models,'' in \emph{Proc. CVPR}, 2022, pp. 10\,684--10\,695.

\bibitem{liu2023audioLDM}
H.~Liu, Z.~Chen, Y.~Yuan, X.~Mei, X.~Liu, D.~Mandic, W.~Wang, and M.~D. Plumbley, ``Audioldm: text-to-audio generation with latent diffusion models,'' in \emph{Proc. ICML}, ser. ICML'23.\hskip 1em plus 0.5em minus 0.4em\relax JMLR.org, 2023.

\bibitem{boll1979suppression}
S.~Boll, ``Suppression of acoustic noise in speech using spectral subtraction,'' \emph{IEEE Transactions on acoustics, speech, and signal processing}, vol.~27, no.~2, pp. 113--120, 1979.

\bibitem{lim1979enhancement}
J.~S. Lim and A.~V. Oppenheim, ``Enhancement and bandwidth compression of noisy speech,'' \emph{Proceedings of the IEEE}, vol.~67, no.~12, pp. 1586--1604, 1979.

\bibitem{ephraim1984speech}
Y.~Ephraim and D.~Malah, ``Speech enhancement using a minimum-mean square error short-time spectral amplitude estimator,'' \emph{IEEE Transactions on acoustics, speech, and signal processing}, vol.~32, no.~6, pp. 1109--1121, 1984.

\bibitem{lu2013speech}
X.~Lu, Y.~Tsao, S.~Matsuda, and C.~Hori, ``Speech enhancement based on deep denoising autoencoder.'' in \emph{Proc. Interspeech}, vol. 2013, 2013, pp. 436--440.

\bibitem{pandey2019new}
A.~Pandey and D.~Wang, ``A new framework for cnn-based speech enhancement in the time domain,'' \emph{IEEE/ACM Transactions on Audio, Speech, and Language Processing}, vol.~27, no.~7, pp. 1179--1188, 2019.

\bibitem{pascual17_interspeech}
S.~Pascual, A.~Bonafonte, and J.~Serrà, ``Segan: Speech enhancement generative adversarial network,'' in \emph{Proc. Interspeech}, 2017, pp. 3642--3646.

\bibitem{leglaive2018variance}
S.~Leglaive, L.~Girin, and R.~Horaud, ``A variance modeling framework based on variational autoencoders for speech enhancement,'' in \emph{2018 IEEE 28th international workshop on machine learning for signal processing (MLSP)}.\hskip 1em plus 0.5em minus 0.4em\relax IEEE, 2018, pp. 1--6.

\bibitem{welker22_interspeech}
S.~Welker, J.~Richter, and T.~Gerkmann, ``Speech enhancement with score-based generative models in the complex stft domain,'' in \emph{Proc. Interspeech}, 2022, pp. 2928--2932.

\bibitem{iwamoto22_interspeech}
K.~Iwamoto, T.~Ochiai, M.~Delcroix, R.~Ikeshita, H.~Sato, S.~Araki, and S.~Katagiri, ``How bad are artifacts?: Analyzing the impact of speech enhancement errors on asr,'' in \emph{Proc. Interspeech}, 2022, pp. 5418--5422.

\bibitem{plchot2016audio}
O.~Plchot, L.~Burget, H.~Aronowitz, and P.~Matejka, ``Audio enhancing with dnn autoencoder for speaker recognition,'' in \emph{Proc. ICASSP}.\hskip 1em plus 0.5em minus 0.4em\relax IEEE, 2016, pp. 5090--5094.

\bibitem{eskimez2018front}
S.~E. Eskimez, P.~Soufleris, Z.~Duan, and W.~Heinzelman, ``Front-end speech enhancement for commercial speaker verification systems,'' \emph{Speech Communication}, vol.~99, pp. 101--113, 2018.

\bibitem{wu2021joint}
Y.~Wu, L.~Wang, K.~A. Lee, M.~Liu, and J.~Dang, ``Joint feature enhancement and speaker recognition with multi-objective task-oriented network.'' in \emph{Proc. Interspeech}, 2021, pp. 1089--1093.

\bibitem{garcia2011analysis}
D.~Garcia-Romero and C.~Y. Espy-Wilson, ``Analysis of i-vector length normalization in speaker recognition systems.'' in \emph{Proc. Interspeech}, vol. 2011, 2011, pp. 249--252.

\bibitem{jung20_odyssey}
J.-W. Jung, J.-H. Kim, H.-J. Shim, S.~bin Kim, and H.-J. Yu, ``Selective deep speaker embedding enhancement for speaker verification,'' in \emph{The Speaker and Language Recognition Workshop (Odyssey 2020)}, 2020, pp. 171--178.

\bibitem{kwon2021ins}
Y.~Kwon, H.-S. Heo, B.-J. Lee, and J.~S. Chung, ``The ins and outs of speaker recognition: lessons from {VoxSRC} 2020,'' in \emph{Proc. ICASSP}.\hskip 1em plus 0.5em minus 0.4em\relax IEEE, 2021, pp. 5809--5813.

\bibitem{desplanques2020ecapa}
B.~Desplanques, J.~Thienpondt, and K.~Demuynck, ``{ECAPA-TDNN}: Emphasized channel attention, propagation and aggregation in tdnn based speaker verification,'' in \emph{Proc. Interspeech}, 2020.

\bibitem{chen2022wavlm}
S.~Chen, C.~Wang, Z.~Chen, Y.~Wu, S.~Liu, Z.~Chen, J.~Li, N.~Kanda, T.~Yoshioka, X.~Xiao \emph{et~al.}, ``Wavlm: Large-scale self-supervised pre-training for full stack speech processing,'' \emph{IEEE Journal of Selected Topics in Signal Processing}, vol.~16, no.~6, pp. 1505--1518, 2022.

\bibitem{defossez20_interspeech}
A.~Défossez, G.~Synnaeve, and Y.~Adi, ``Real time speech enhancement in the waveform domain,'' in \emph{Proc. Interspeech}, 2020, pp. 3291--3295.

\bibitem{li2024return}
T.~Li, D.~Katabi, and K.~He, ``Return of unconditional generation: A self-supervised representation generation method,'' in \emph{The Thirty-eighth Annual Conference on Neural Information Processing Systems}, 2024.

\bibitem{lei2016layer}
J.~Lei~Ba, J.~R. Kiros, and G.~E. Hinton, ``Layer normalization,'' \emph{ArXiv e-prints}, pp. arXiv--1607, 2016.

\bibitem{elfwing2018sigmoid}
S.~Elfwing, E.~Uchibe, and K.~Doya, ``Sigmoid-weighted linear units for neural network function approximation in reinforcement learning,'' \emph{Neural networks}, vol. 107, pp. 3--11, 2018.

\bibitem{koizumi23_interspeech}
Y.~Koizumi, H.~Zen, S.~Karita, Y.~Ding, K.~Yatabe, N.~Morioka, M.~Bacchiani, Y.~Zhang, W.~Han, and A.~Bapna, ``Libritts-r: A restored multi-speaker text-to-speech corpus,'' in \emph{Proc. Interspeech}, 2023, pp. 5496--5500.

\bibitem{kahn2020libri}
J.~Kahn, M.~Riviere, W.~Zheng, E.~Kharitonov, Q.~Xu, P.-E. Mazar{\'e}, J.~Karadayi, V.~Liptchinsky, R.~Collobert, C.~Fuegen \emph{et~al.}, ``Libri-light: A benchmark for asr with limited or no supervision,'' in \emph{Proc. ICASSP}.\hskip 1em plus 0.5em minus 0.4em\relax IEEE, 2020, pp. 7669--7673.

\bibitem{huh2024vox}
J.~Huh, J.~S. Chung, A.~Nagrani, A.~Brown, J.-w. Jung, D.~Garcia-Romero, and A.~Zisserman, ``The vox celeb speaker recognition challenge: A retrospective,'' \emph{IEEE/ACM Trans. on Audio, Speech, and Language Processing}, 2024.

\bibitem{heo2023rethinking}
H.-S. Heo, K.~Nam, B.-J. Lee, Y.~Kwon, M.~Lee, Y.~J. Kim, and J.~S. Chung, ``Rethinking session variability: Leveraging session embeddings for session robustness in speaker verification,'' in \emph{Proc. ICASSP}, 2023.

\bibitem{paszke2019pytorch}
A.~Paszke, S.~Gross, F.~Massa, A.~Lerer, J.~Bradbury, G.~Chanan, T.~Killeen, Z.~Lin, N.~Gimelshein, L.~Antiga \emph{et~al.}, ``Pytorch: An imperative style, high-performance deep learning library,'' in \emph{Proc. NeurIPS}, vol.~32, 2019.

\bibitem{loshchilov2018decoupled}
I.~Loshchilov and F.~Hutter, ``Decoupled weight decay regularization,'' in \emph{Proc. ICLR}, 2019.

\bibitem{omid19nsre}
O.~Sadjadi, C.~Greenberg, E.~Singer, D.~Reynolds, L.~Mason, and J.~Hernandez-Cordero, ``The 2018 nist speaker recognition evaluation,'' in \emph{Proc. Interspeech}, 2019.

\end{thebibliography}

\end{document}